\documentclass[sn-mathphys]{sn-jnl}

\jyear{2021}%

\begin{document}

\title[Article Title]{SAMPL9 blind predictions using nonequilibrium alchemical approaches}


\author*[1]{\fnm{Piero} \sur{Procacci}}\email{piero.procacci@unifi.it}

\author[2]{\fnm{Guido} \sur{Guarnieri}}\email{guido.guarniri@enea.it}
\affil*[1]{\orgdiv{Chemistry}, \orgname{University of Florence},
  \orgaddress{\street{Via Lastruccia 3}, \city{Sesto Fiorentino (FI)}, \postcode{50019}, \country{Italy}}}

\affil[2]{\orgdiv{DTE-ITC}, \orgname{ENEA, Portici research center},
  \orgaddress{\street{P.le  E. Fermi 1}, \city{Portici (NA)}, \postcode{80055}, \country{Italy}}}


\abstract{We present our blind predictions for the Statistical Assessment of the
Modeling of Proteins and Ligands (SAMPL), 9th challenge, focusing on
binding of WP6 (carboxy-pillar[6]arene) with ammonium/diammonium
cationic guests. Host-guest binding free energies have been calculated
using the recently developed virtual double system single box
approach, based on the enhanced sampling of the bound and unbound
end-states followed by fast switching nonequilibrium alchemical
simulations [M Macchiagodena, M Pagliai, M Karrenbrock, G Guarnieri, F
  Iannone, Procacci, J Chem Theory Comput, 16, 7260, 2020].  As far as
Pearson and Kendall coefficients are concerned,
performances were acceptable and, in general, better than those we
submitted for calixarenes, cucurbituril-like open cavitand, and
beta-cyclodextrines in previous SAMPL host-guest challenges, confirming
the reliability of nonequilibrium approaches for absolute binding free
energy calculations.  In comparison with previous submissions, we found
nonetheless a rather large mean signed error that we attribute to the
way the finite charge correction was addressed through the
assumption of a neutralizing background plasma.
}




\maketitle

\section {Introduction}
The Statistical Assessment of the Modeling of Proteins and Ligands
(SAMPL) is a set of community-wide blind challenges aimed at advancing
the computational techniques in drug
design.\cite{Muddana2014,Yin2016,Rizzi2018,Rizzi2019,Amezcua2021}. New
experimental data, such as dissociation free energies, hydration free
energies, acid-base dissociation constants, or partition coefficients
are withheld from participants until the prediction submission
deadline, so that the true predictive power of methods can be
assessed.
\begin{figure}[h]
  \begin{center}
    \includegraphics[scale=0.18]{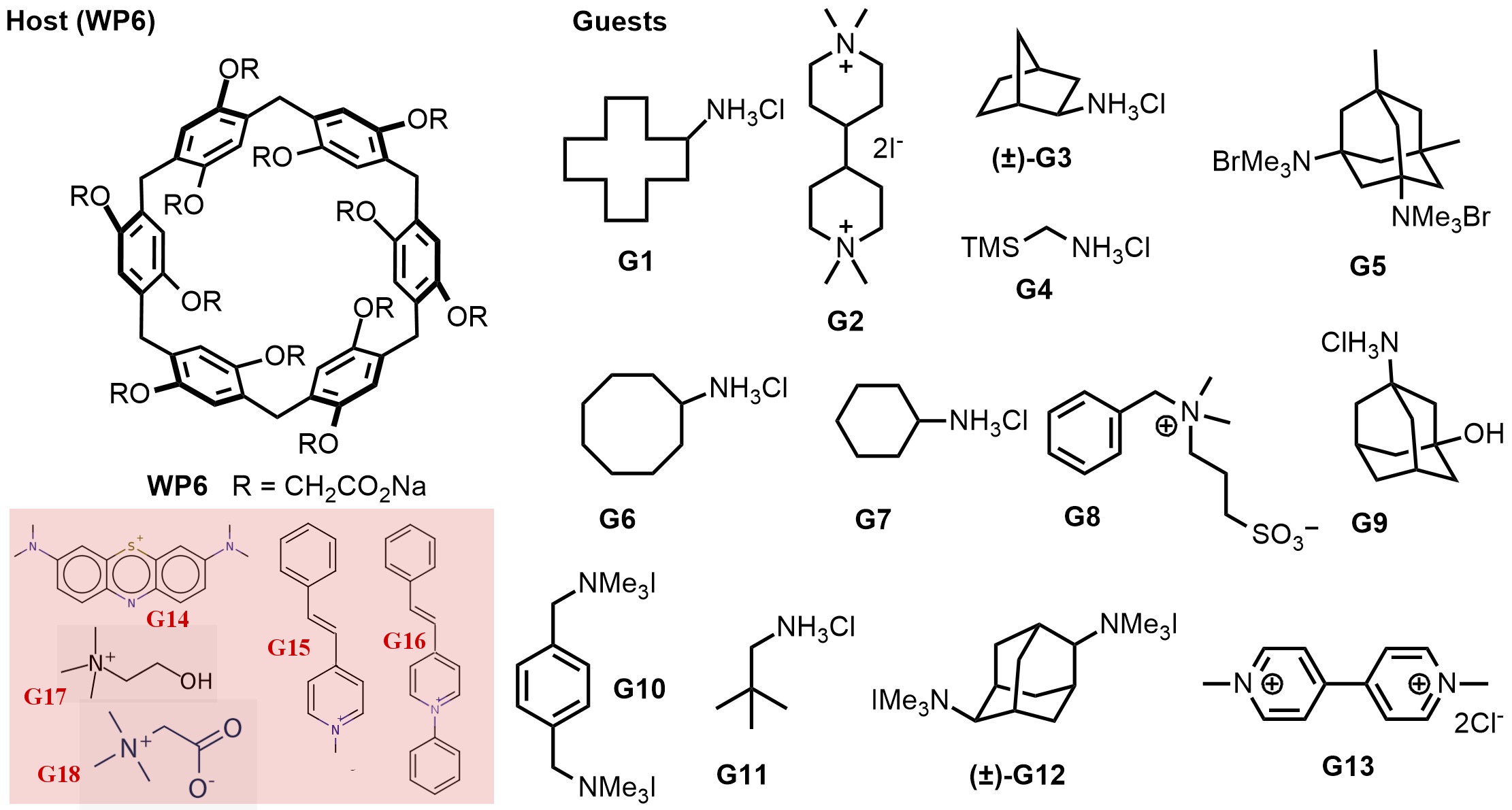}
    \caption{Host-guest SAMPL9 challenge. In the shaded red area on
      the bottom left corner, the
      molecules with known binding affinities used for pre-assessment are reported (see text). }
  \end{center}
  \label{fig:sampl9}
\end{figure}

In the latest 9th challenge, participants were required to predict the
binding affinities of thirteen ammonium/diammonium cationic ligands
(guests) vs WP6, a water-soluble toroidal macrocyclic host molecule
(see Figure \ref{fig:sampl9}). Host-guest experimental binding
affinities, measured by Isothermal Titration Calorimetry (ICT) at
pH=7.4 were disclosed by November 2021 and are now reported in
Ref. \cite{sampl9exp}.  Actually, the binding affinity of G13
(Paraquat) was available on literature at physiological
pH\cite{Yu2012,Nicolas2017} {\it before} the submission deadline.  The
host WP6, never featured in previous SAMPL challenges, is structurally
similar to cucurbit[n]urils (CBn) used in SAMPL6\cite{Rizzi2018} and
SAMPL8\cite{SAMPL8} with important differences related to the
electrostatic interactions.  Unlike the neutral CBn host, the
quasi-D6h WP6 bears six carboxylated moieties on the upper and lower
rims that can be in part protonated at
pH=7.4.\cite{Yu2012,Nicolas2017} As the pKa's of WP6 wre not
available, such a feature represented an important challenge in the
blind prediction of WP6-guest binding affinities.  Participants were
in fact requested to deal with this complication.

In this report, we present the dissociation free energies of the fully
anionic WP6$^{-12}$ for the thirteen cationic guests of Figure 1 plus
five additional guest molecules with known dissociation constants,
namely methylene blue (G14)\cite{Yang2018}, GNF-Pf-3194
(G15)\cite{Hessz2021}, M2 (G16)\cite{Hessz2021}, choline
(G17)\cite{Hua2018} and betaine (G18)\cite{Hua2018}.  The calculations
have been performed on the CRESCO6-ENEA cluster\cite{cresco} in
Portici (Italy) using the so-called virtual Double System Single Box
(vDSSB) method\cite{hcq,vdssb,springprot,vdssbg}, based on a
production of a swarm of concurrent nonequilibrium (NE) alchemical
simulations that are started from end-state canonical configurations
sampled using Hamiltonian Replica Exchange (HREM)\cite{Okamoto1999}.

\section{Methods}

\subsection{Estimation of the protonation state of WP6}
As stated in the introduction, the twelve pKa's of WP6 host were not
known before the SAMPL9 deadline for submission. In presence of
multiple host-guest complexes with various protonation states,
the overall observed association constant
for Gn-WP6 complexes is given by
\begin{equation}
  K_a = \sum_k W_k K_a^{(k)}
  \label{eq:ka}
  \end{equation}
where $K_a^{(k)}$is the association constant for the $k$-th protonated
WP6 species and $W_k$ is the corresponding normalized weight. The Gn-WP6
dissociation free energy should be hence calculated as
\begin{equation}
  \Delta G_d = RT \ln \sum_k W_k e^{\beta \Delta G_d^{(k)}}
  \label{eq:DGk}
\end{equation}
where $\Delta G_d^{(k)} $is the dissociation free energy with the host
in the $k$-th protonation state. The predicted pKa of the WP6 template
monobasic acid 2-(2,5-dimethylphenoxy) acetic acid is
3.23\cite{scifind}.  We assumed a pKa distribution modulated by the so-called statistical
factor\cite{statfac} for twelve equivalent protonation sites
\begin{equation}
  \rm{pKa}_n = \rm{pKa}_1-\log(n/13-n) 
\end{equation}
where ${\rm pKa}_1=3.23$ is that of the monobasic acid.
We obtained that the prevalent species at pH=7.4 is the WP6$^{-12}$
anion with all deprotonated carboxylated groups with a weight
$W_{-12}=0.98$, in surprisingly good agreement, despite the crudeness
of our approach, with the experimental value of $W_{-12}=0.95$
released after the submission deadline.\cite{SAMPL9}

We hence computed the dissociation constant for the WP6$^{-12}$
species only,  related to the chemical equilibrium
\begin{equation}
  \rm {WP6^{-12}Gn } \leftrightarrow \rm{WP6^{-12} + Gn}.
  \label{eq:anion}
\end{equation}
It should be
nonetheless stressed that the large weight of the fully depronotated
species, W=0.95, does not guarantee that the experimental dissociation
free energy corresponds to the dissociation
free energy of the equilibrium Eq. \ref{eq:anion}. If the dissociation
free energy of any of the other simultaneous chemical equilibria,
$\rm {WP6^{-n}Gn } \leftrightarrow \rm{WP6^{-n} + Gn}$ with
$n=1 ... 11$, 
is much
{\it larger} than that corresponding to the equilibrium
Eq. \ref{eq:anion}, then, according to Eq. \ref{eq:DGk}, the overall
experimental free energy can be significantly higher than that
matching the equilibrium Eq. \ref{eq:anion}. So our dissociation free
energy, computed for the equilibrium \ref{eq:anion}, should be viewed
as a {\it lower bound} for the true experimental dissociation free
energy.

\subsection{vDSSB approach for absolute dissociation free energies} 
The vDSSB methodology is thoroughly described in
Ref. \cite{vdssb,springprot}.  In brief, the method consists of two
massively parallel computational steps, the HREM stage and the
nonequilibrium alchemical stage. The HREM stage relies on the enhanced
sampling of the host-guest bound state (with the ligand at full
coupling) in explicit water and of the isolated guest molecule. The
initial configurations for the unbound state are prepared by combining
the HREM-sampled gas-phase (decoupled) ligand  snapshots with a
pre-equilibrated box filled with explicit water.  HREM uses $n=16$ and
$n=8$ replicas for the bound and unbound states respectively, with a
maximum scaling factor of $S=0.1$ (corresponding to a temperature of
3000 K) involving only the intra-solute bonded and non-bonded
interactions\cite{Marsili2010}.  The scaling factors along the replica
progression are computed according to the protocol $S_m = S^{(m-1)/n}$
with $m=1...n$.  Only the scaling factors are exchanged among
neighboring replicas to minimize the communication overhead on the MPI
layer. The HREM simulations are run for 48 ns and 16 ns for the bound
and gas-phase state, respectively with exchange rates in the range
25\%-50\%.

In the NE alchemical stage, the bound state leg of the
alchemical cycle is performed by rapidly decoupling the bound ligand
in a swarm of 360 independent NE simulations, each lasting 1.44
ns. The unbound leg of the cycle is done by growing (recoupling) the
ghost ligand in the solvent in 480 NE alchemical simulations lasting
0.36 ns. Electrostatic and Lennard-Jones interactions are switched
off/on sequentially, in order to separate the corresponding
contributions to the binding. For further details on the ligand
coupling/decoupling protocols, we refer to
Ref. \cite{vdssb,springprot}.

The final bound and unbound
alchemical work distributions are combined in the convolution
\begin{equation}
  P(W) = P_b\ast P_u(W)= \int dw P_b(W)P(W-w)dw
  \label{eq:conv}
\end{equation}
as if the two independent processes of the ligand annihilation in the
bound state and of the ligand growth in the pure solvent occurred in the same
box (hence the name  vDSSB), with one ligand on the host and the other in the far distant bulk. 

The dissociation free estimate can be performed by way of the
Jarzynski identity\cite{Jarzynski1997}
\begin{equation}
  \Delta G_{\rm vDSSB} = -
  RT  \ln \int P(W) e^{-\beta W}  dW  \\
  \label{eq:jarz}
\end{equation}  
 or using the Crooks theorem\cite{Crooks1998} in the assumption that
 the convolution can be described by a mixture of normal
 distributions\cite{vdssb,Procacci2015}, i.e
\begin{equation}
  \Delta G_{\rm vDSSB} = -
  RT  \ln \sum_i c_i e^{-\beta(\mu_i -\frac{1}{2}\beta\sigma_i^2)}
  \label{eq:em}
\end{equation}  
where $c_i,~\mu_i,~\sigma_i^2$ are the normalized weight, mean and
variance of the $i$-th normal component, determined via the
expectation-maximization algorithm \cite{emorac,em}.  Estimates based
on eq. \ref{eq:jarz} or eq. \ref{eq:em} are used depending on the
width and character of the work distributions as assessed by the
Anderson-Darling normality test \cite{Anderson1954}. The free energy
estimates are corrected for a volume\cite{pccp1,procsampl6,procsampl7}
and a charge term\cite{Darden1998,procsampl6} (for non-neutral ligand)
given by
\begin{equation}
  \Delta G_{\rm vol} = RT \ln \left [ 4\pi \frac{(2\sigma)^3}{3V_0}
  \right ]
  \label{eq:dgvol} 
\end{equation}
\begin{equation}
\Delta G_{\rm fs} =  - \frac{\pi}{2\alpha^2} \left \{ \frac{ [ Q_H^2  - (Q_H+Q_G)^2 ] }{
  V_{\rm BOX}^{(b)}}   +  \frac {Q_G^2 }{V_{\rm BOX}^{(u)}} \right \}  
\label{eq:fs}
\end{equation}
where $\sigma$ is the standard deviation of the host-guest COM-COM
distance distribution in the bound state, $V_0$ is the standard state
volume, $Q_G$, $Q_H$ are the net charges of the guest and the
host, $V_{\rm BOX}^{(b)}$, $V_{\rm BOX}^{(u)}$ and are the mean volume
of the MD box for the bound and unbound state respectively, and
$\alpha$ is the Ewald convergence parameter.  The effectiveness of the
correction Eq. \ref{eq:fs} for finite size effects when dealing with
charged ligands has been amply
assessed in \cite{procsampl6}, while the origin of the binding volume
term Eq. \ref{eq:dgvol} has been discussed in
Refs. \cite{pccp1,procacci2018,procsampl7}. The final blind prediction
for the standard dissociation free energy is given by
\begin{equation}
  \Delta G_d^0 = \Delta G_{\rm vDSSB} + \Delta G_{\rm vol} + \Delta G_{\rm
    fs}
  \label{eq:DG0}
\end{equation}
\subsection{Simulation setup and parameters}
Guests and host PDB files were taken from the SAMPL9 GitHub site.  For
the two chiral compounds, the G3 guest
(2-bicyclo[2.2.1]heptanyl]azanium) corresponds to the 1R 2R 4S
  diastereoisomer. The G12 guest (Hexamethyladamantane-2,6-diammonium)
  has the 2 and 6 carbon atoms of the R type.  The Force Field (FF)
  parameters and topology of the host and guests molecules were
  prepared using the PrimaDORAC interface\cite{primadorac} based on
  the GAFF2\cite{gaff2} parameter set. For G4, the silicon-related
  parameters were taken from Ref. \cite{Dong2021}.  The initial bound
  state was prepared using the Autodock Vina code\cite{vina}. The
  bound complexes and the ghost ligands were solvated in about 1600
  and 512 OPC3\cite{opc3} water molecules, respectively. Long-range
  electrostatic interactions were treated using the Smooth Particle
  Mesh Ewald method (SPME)\cite{Essmann1995}. A background
  neutralizing plasma was assumed within the SPME
  method\cite{Darden1998}. The cut-off of the Lennard-Jones
  interactions was set to 13 \AA. All simulations, HREM or
  nonequilibrium, were performed in the NPT ensemble in standard
  conditions using an isotropic Parrinello-Rahman
  Langrangian\cite{Parrinello1980} and a series of Nos\'e
  thermostats\cite{Nose1984} for pressure and temperature control,
  respectively. Bonds constraints were imposed on X-H bonds only,
  where X is a heavy atom. All other bonds were assumed to be
  flexible. In the bound state simulations (HREM or NE) the guest and
  the host center of mass (COM) were tethered by a weak harmonic
  potential with a force constant of $K=0.06$ kcal/mol $\AA^{-2}$,
  corresponding to a ligand allowance volume of $\simeq 600 \AA^3$.
  All simulations have been done using the hybrid OpenMP-MPI program
  ORAC\cite{orac6} on the CRESCO6 cluster\cite{cresco}.
\section{Results}
\begin{figure}[h]
  \begin{center}
  \includegraphics[scale=0.45]{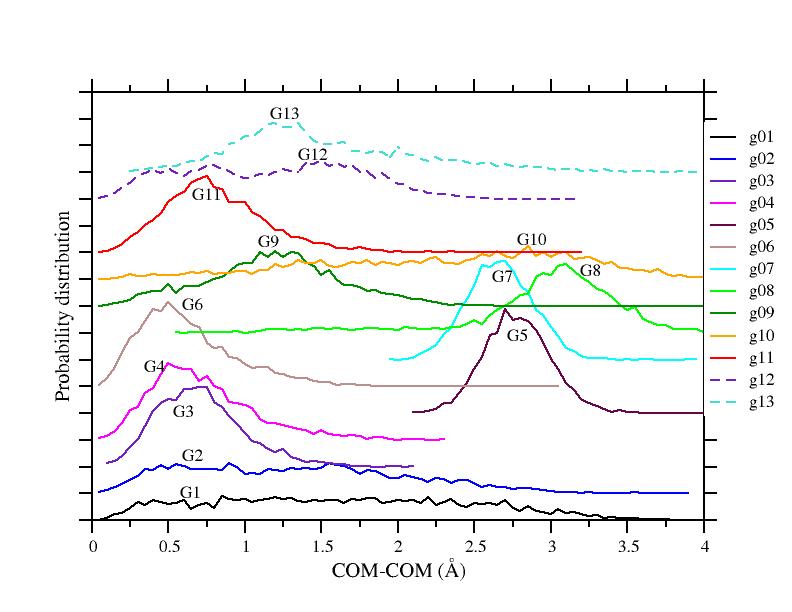}
  \caption{Host-guest COM-COM distance distributions obtained
    from the HREM sampling of the bound state.}
  \end{center}
    \label{fig:com}
\end{figure}

In Figure \ref{fig:com} we show the host-guest COM-COM distance
distributions in the bound state for all thirteen SAMPL9 pairs as
obtained from the target state HREM sampling of 48 ns.  According to
Eq. \ref{eq:dgvol}, the binding site volume is calculated from the
variance of the COM-COM distribution as $V_{\rm site} = 4\pi
\frac{(2\sigma)^3}{3}$.
As it can be seen, despite the weak tethering COM-COM harmonic
potential and the correspondingly large allowance radius of $\simeq$8
\AA~(exceeding substantially the radius of the WP6 cavity of
$\simeq$5.4 \AA), the guests never leave the WP6 pocket, with a
standard deviation of the COM-COM distribution ranging from
$\simeq$0.2 \AA~ (G7) to $\simeq$ 0.9 \AA~ (G10). As discussed in
Ref. \cite{procsampl7}, the HREM approach with {\it intrasolute}
(ligand and host) scaling does not accelerate passive diffusion with
little impact on the $k_{\rm off}$ and on the ligand residence time
(in the range of the microsecond to milliseconds for all ligands).  On
the Zenodo platform (https://zenodo.org/record/5891191) we provide the
HREM-generated trajectories of the bound state for all thirteen
guest-host systems.

In Table \ref{tab:dat} we have collected all the quantities that
have been used in the determination of the estimate of the standard
dissociation free energy, Eq. \ref{eq:DG0}, namely the charge and
volume corrections (Eq. \ref{eq:dgvol} and Eq. \ref{eq:fs}), the mean
and variance of the bound and unbound work distributions as well as
their character as assessed by the Anderson-Darling test.  The bound
and unbound work distributions, along with their convolution
Eq. \ref{eq:conv}, are reported for all guests of Table \ref{tab:dat}
in the Electronic Supporting Information (ESI). The corresponding raw
data of the work can be found on the Zenodo site (https://zenodo.org/record/5891191).
\begin{table}[h]
  \begin{tabular}{ccccccccccl}
    \hline
    guest & $\Delta G_{\rm vol}$ & $\Delta G_{\rm fs}$ & $\langle W_b
    \rangle $  & $\sigma_b$ & $\langle W_u \rangle $  & $\sigma_u$ &
    ${\rm AD}_b$ & ${\rm AD}_u$ & $\sigma_{BU}$ & Est. type\\
    \hline
 G1  &  -2.6  &  -2.0 &   75.9 & 3.2 &  -53.0 & 1.9 &  2.32  &  2.96 &  3.73 & $S_{J}+B$ \\
 G2  &  -2.9  &  -4.3 &  147.6 & 2.3 & -121.2 & 1.1 &  0.41  &  0.24 &  2.53 &  $S_3$ \\
 G3  &  -4.5  &  -2.0 &   78.5 & 2.8 &  -56.8 & 0.8 &  2.57  &  0.63 &  2.88 &  $S_3$ \\
 G4  &  -4.2  &  -2.0 &   73.8 & 2.2 &  -56.8 & 0.8 &  0.38  &  0.31 &  2.35 &   $G_b$ \\
 G5  &  -4.5  &  -4.3 &  144.0 & 2.3 & -122.6 & 1.3 &  0.50  &  0.29 &  2.61 &  $S_3$ \\
 G6  &  -4.3  &  -2.0 &   78.6 & 2.9 &  -55.4 & 0.9 &  1.34  &  0.45 &  3.02 & $S_{J}+B$ \\
 G7  &  -5.0  &  -2.0 &   82.6 & 3.0 &  -57.0 & 0.8 &  0.46  &  0.90 &  3.14 &  $S_3$ \\
 G8  &  -3.3  &   0.0 &   75.5 & 3.2 &  -49.9 & 1.6 &  1.34  &  0.50 &  3.57 &  $S_3$ \\
 G9  &  -3.6  &  -2.0 &   83.1 & 2.8 &  -59.4 & 1.0 &  0.67  &  0.47 &  2.99 & $S_{J}+B$ \\
 G10  &  -2.6  &  -4.3 &  143.7 & 2.3 & -119.5 & 1.1 &  0.65  &  0.23 &  2.51 &  $S_3$ \\
 G11  &  -4.2  &  -2.0 &   76.3 & 2.3 &  -58.5 & 0.8 &  0.81  &  0.70 &  2.41 &  $S_3$ \\
 G12  &  -3.4  &  -4.3 &  151.3 & 2.1 & -122.1 & 1.1 &  0.25  &  0.22 &  2.37 &   $G_b$ \\
 G13  &  -3.0  &  -4.3 &  144.5 & 2.4 & -125.4 & 0.9 &  0.28  &  0.18 &  2.60 &   $G_b$ \\
  \hline\hline
 G14  &  -2.8  &  -2.0 &   44.5 & 1.9 &  -21.2 & 1.1 &  0.30  &  0.31 &  2.18  &   $G_b$ \\
 G15  &  -2.8  &  -2.0 &   51.7 & 2.1 &  -32.1 & 0.9 &  0.42  &  0.29 &  2.31  &  $S_3$ \\
 G16  &  -2.7  &  -2.0 &   49.1 & 2.0 &  -27.5 & 1.1 &  0.73  &  0.40 &  2.26  &  $S_3$ \\
 G17  &  -4.3  &  -2.0 &   57.3 & 1.6 &  -40.9 & 0.7 &  0.45  &  0.36 &  1.72  &  $S_3$ \\
 G18  &  -3.7  &   0.0 &   44.5 & 1.7 &  -33.4 & 0.9 &  1.22  &  0.52 &  1.91  & $S_{J}+B$ \\ \hline
\end{tabular}
\caption{Salient data for vDSSB predictions. $\Delta G_{\rm
    vol}$, standard state correction; $\Delta G_{\rm fs}$, finite size
  correction for charged guest molecule; $\langle W_{b/u} \rangle $,
$\sigma_{b/u}$, ${\rm AD}_b$,  mean NE work, standard deviation and
  Anderson Darling test for the bound and unbound states;
  $\sigma_{BU}$,  standard deviation of the convolution;   Est. type,
  estimate type (see text). Energy unit is in kcal/mol. }
\label{tab:dat}
\end{table}
\noindent
The last column of Table \ref{tab:dat} indicates the type of estimate
for the alchemical dissociation free energies $\Delta G_{\rm vDSSB}$
of the ranked submission: $G_b$ refers to a Gaussian estimate, $S_3$
is an estimate based on the Gaussian mixture, Eq. \ref{eq:em}, with
three components, and finally the estimate $S_J+B$ is based on the
Jarzynsky exponential average, Eq. \ref{eq:jarz}.

Ideally, such estimates should all converge to the same value when $n
\rightarrow \infty$ and/or the duration of the NE process $\tau
\rightarrow \infty$.  For finite samples, their precision and accuracy
are strongly affected by the spread (connected to the dissipation of
the process\cite{procsampl7}) and the character of the work distributions.
The criteria for selecting the estimate type for each pair were the
following: i) if both the bound and unbound work distributions prior
to the convolution pass the Anderson-Darling test with a
p-value\cite{pvalue} greater than 50\% ($A^2<0.3$), then the Gaussian
estimate is adopted; otherwise, ii) if the uncertainty is less than 2
kcal/mol we use $S_3$; otherwise iii) we use the Jarzynski estimate
$S_J+B$, where to Eq. \ref{eq:jarz} is added a correction due to the
bias.  We recall that the {\it positive bias} bias in exponential
averaging of the NE work is inescapable\cite{Gore2003,Procacci2021}.
\begin{figure}[h]
  \begin{center}
  \includegraphics[scale=0.35]{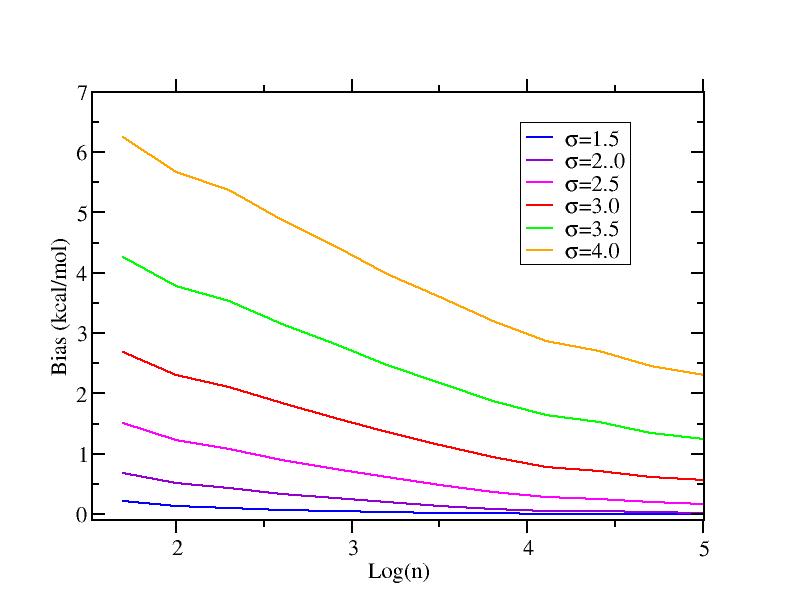}
  \caption{Bias of the exponential average as a function of n and $\sigma$}
  \end{center}
    \label{fig:bias}
\end{figure}
The convolution, by boosting the sample size to $n=n_b\times n_u$,
tames the bias but does not suppress it, notably when the width of the
resulting distribution is large. As one can see from Table
\ref{tab:dat} and Figures S1-S18 of the ESI, in general, the width of
the convolution is somewhat larger than the width of the {\it bound
  state} distribution, since $\sigma^2_u < \sigma_b^2$, even if the
duration time of the ligand annihilation in the bound state process
(1.44 ns) is much longer than that of the ligand growth process in the bulk
(0.36 ns).  The variance of the convolution is in fact given by
$\sigma_{BU}^2\simeq \sigma_b^2+\sigma_u^2$ (an equation that is {\it
  exact} if both distributions are normal).  From the variance of the
convolution $\sigma^2_{BU}$, the bias can be readily estimated by a
generating a sample of {\it normal distributions} with zero mean and variance
$\sigma^2_{BU}$ and by computing the bias as
\begin{equation}
  B(\sigma,n) =   \left \langle \frac{1}{n} \sum^n e^{-\beta W} \right \rangle + \frac{1}{2}\sigma^2_{bu}
  \label{eq:bias}
\end{equation}
where the exponential sum is averaged over all generated normal
distributions.  In Figure \ref{fig:bias}, we report the bias,
Eq. \ref{eq:bias} as a function of the number of points and of the
variance. In our case, $n=360\times480=182400$ while $\sigma_{bu}$
varies in the range 2.3 (G4) up to 3.7 (G1), with a bias correction
of the order of the fraction of kcal/mol in most cases and slightly
higher than 1 kcal/mol only for G6 , G7, G8 and G1.

In Table \ref{tab:dgs}, we report in the second column the vDSSB
submitted blind predictions (where the estimate was determined
according to the previously cited criteria) along with the Gaussian
estimate, the Gaussian mixture estimate (Eq. \ref{eq:em}), and the
bias-corrected Jarzynski estimate (not submitted). The uncertainties have been
computed by bootstrapping with resampling the bound and unbound work
samples {\it prior to} convolution, to avoid an artificial
underestimation of the error when resampling from a boosted
distribution\cite{vdssb}. 

\begin{table}[h]
  \begin{tabular}{rrrrrrr}
    \hline\hline
    guest & Pred. & Exp.  & Gauss  & EM(3) & Jarz+bias & $\Delta G_q$
    \\
    \hline
 G1 &    7.8$\pm$  0.6   &{\bf  -6.53  } &     6.7 $\pm$ 1.1  &   8.5$\pm$  2.6  &   7.8$\pm$  0.6  &  -0.7 $\pm$ 1.2 \\
 G2 &   13.0$\pm$  1.2   &{\bf -10.59  } &    13.7 $\pm$ 0.8  &  13.0$\pm$  1.2  &  14.0$\pm$  0.6  &   1.0 $\pm$ 0.3 \\
 G3 &   11.0$\pm$  0.8   &{\bf  -8.03  } &     8.2 $\pm$ 0.6  &  11.0$\pm$  0.8  &  11.1$\pm$  0.3  &   0.5 $\pm$ 0.8 \\
 G4 &    6.2$\pm$  0.8   &{\bf  -6.50  } &     6.2 $\pm$ 0.4  &   7.3$\pm$  0.4  &   7.2$\pm$  0.4  &   1.2 $\pm$ 0.4 \\
 G5 &    7.2$\pm$  0.6   &{\bf  -5.46  } &     6.8 $\pm$ 0.6  &   7.2$\pm$  0.6  &   7.4$\pm$  0.5  &   5.3 $\pm$ 0.2 \\
 G6 &   11.6$\pm$  0.4   &{\bf  -8.08  } &     9.2 $\pm$ 1.2  &  11.9$\pm$  1.2  &  11.6$\pm$  0.4  &   0.3 $\pm$ 0.8 \\
 G7 &   11.5$\pm$  1.9   &{\bf  -7.07  } &    10.3 $\pm$ 0.9  &  11.5$\pm$  1.9  &  12.3$\pm$  0.6  &  11.1 $\pm$ 0.7 \\
 G8 &   11.6$\pm$  1.6   &{\bf  -6.04  } &    11.5 $\pm$ 1.1  &  11.6$\pm$  1.6  &  12.1$\pm$  0.8  &   2.1 $\pm$ 1.5 \\
 G9 &   11.8$\pm$  0.9   &{\bf  -6.32  } &    10.4 $\pm$ 1.0  &  12.2$\pm$  1.9  &  11.8$\pm$  0.9  &  -0.5 $\pm$ 0.8 \\
 G10 &   12.7$\pm$  0.7   &{\bf  -9.96  } &    12.0 $\pm$ 0.5  &  12.7$\pm$  0.7  &  13.0$\pm$  0.5  &   1.2 $\pm$ 0.3 \\
 G11 &    8.3$\pm$  0.5   &{\bf  -6.26  } &     6.7 $\pm$ 0.4  &   8.3$\pm$  0.5  &   8.4$\pm$  0.4  &   0.7 $\pm$ 0.5 \\
 G12 &   16.8$\pm$  0.8   &{\bf -11.02  } &    16.8 $\pm$ 0.4  &  17.5$\pm$  0.3  &  17.6$\pm$  0.3  &   2.8 $\pm$ 0.2 \\
 G13 &    6.9$\pm$  1.0   &{\bf  -8.58  } &     6.1 $\pm$ 0.8  &   6.9$\pm$  0.8  &   6.9$\pm$  0.5  &   1.5 $\pm$ 0.5 \\
    \hline\hline
 G14 &   14.5$\pm$  0.7   &{\bf   9.68  } &    14.5 $\pm$ 0.7  &  15.0$\pm$  0.8  &  15.0$\pm$  0.6  &   5.2 $\pm$ 0.1 \\
 G15 &   10.7$\pm$  0.6   &{\bf   8.37  } &    10.4 $\pm$ 0.5  &  10.7$\pm$  0.6  &  10.8$\pm$  0.6  &   4.9 $\pm$ 0.2 \\
 G16 &   12.9$\pm$  0.4   &{\bf  10.59  } &    12.7 $\pm$ 0.6  &  12.9$\pm$  0.4  &  13.0$\pm$  0.3  &   5.1 $\pm$ 0.2 \\
 G17 &    8.0$\pm$  0.2   &{\bf   6.48  } &     7.5 $\pm$ 0.3  &   8.0$\pm$  0.2  &   8.1$\pm$  0.2  &   2.5 $\pm$ 0.2 \\
 G18 &    4.9$\pm$  0.4   &{\bf   0.00  } &     4.4 $\pm$ 0.5  &   4.9$\pm$  0.4  &   4.9$\pm$  0.4  &  -3.8 $\pm$ 0.3 \\
    \hline
  \end{tabular}
\caption{Estimates of the dissociation free energy in the vDSSB
  approach. Pred., submitted predictions; Exp., ITC data; Gauss,
  Gaussian estimates; EM(3), estimates based on Eq. \ref{eq:em};
  Jarz+bias, estimates based on Eq. \ref{eq:jarz}; $\Delta G_q$,
  estimated contribution of the electrostatic interactions to the
  binding (computed using the Gaussian assumption).  Energy units are
  in kcal/mol. }
  \label{tab:dgs}
\end{table}
\noindent
In the last column, we report the electrostatic contribution ($\Delta
G_q$) to the
overall dissociation free energy. In general, we can say that
electrostatics is not the driving force for binding. In two cases, the
electrostatic balance is unfavorable (G9, G18). For the small
G7, most of the binding comes instead from electrostatics, very
likely due to the presence of H-bonded water molecules at the entrance
of the binding pocket as can be verified from the HREM trajectories
provided on Zenodo (https://zenodo.org/record/5891191).
\begin{figure}[h]
  \begin{center}
  \includegraphics[scale=0.40]{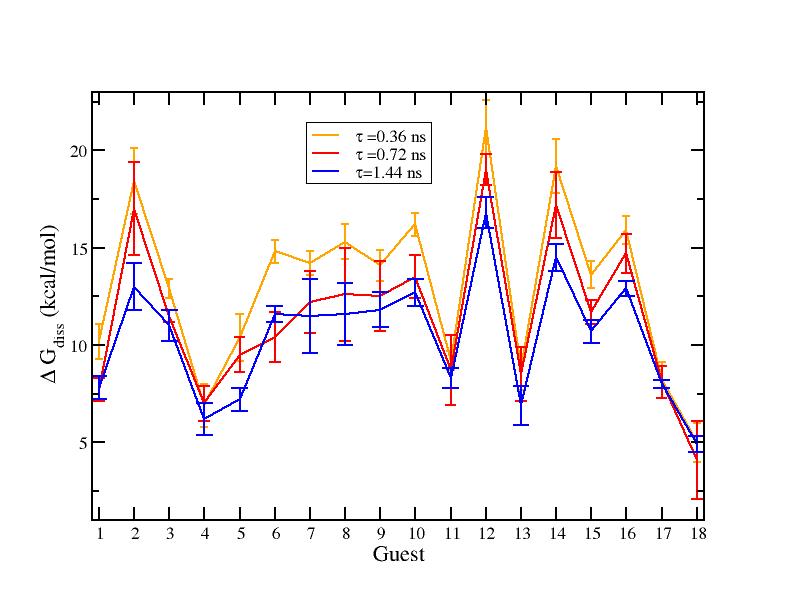}
  \end{center}
  \caption{Convergence of vDSSB estimates of the dissociation free
    energy as a function of the duration time of the
  bound state NE alchemical transitions for the eighteen guest
  molecules reported in Figure \ref{fig:sampl9}.}
    \label{fig:conv}
\end{figure}
The convergence of the prediction can be assessed by analyzing the vDSSB estimates,
based on the above criteria, as a function of the duration time of
the most dissipative process, i.e. the annihilation of the ligand in
the bound state  (see $\sigma_u$ and $\sigma_b$ entries in Table
\ref{tab:dat}).  From Figure \ref{fig:conv}, we can see that most of
the predictions appear to have converged within the uncertainty,
with only two exceptions (G2 and G5).

\begin{figure}[h]
  \begin{center}
  \includegraphics[scale=0.40]{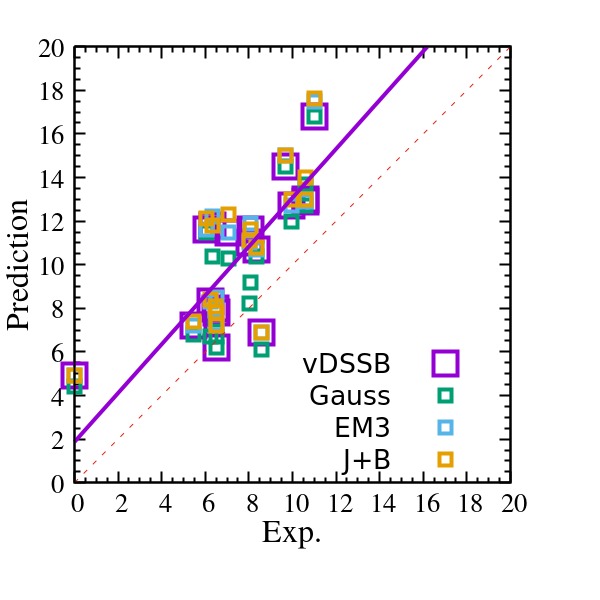}
  \end{center}
  \caption{Correlation diagram for vDSSB-based dissociation free
    energy estimates reported in
    Table \ref{tab:dgs}}
    \label{fig:corr}
\end{figure}

The correlation diagrams and metrics for all the estimates of Table
\ref{tab:dgs} are reported in Figure \ref{fig:corr} and Table \ref{tab:metr}. The large squares refer to the vDSSB ranked
prediction. All predictions based on the four estimates are
strongly correlated one to the other, whether one chooses the sets
including only thirteen guests from SAMPL9, the five additional ligands, and the
set comprising all the eighteen ligands.  Regarding the agreement with ITC
data, we note that, while the Pearson's and Kendall correlation
\noindent
coefficients $R_{xy}$ are acceptable (for the 13 SAMPL9 guests), good (for all
the 18 systems) and excellent (for the 5 extra ligands), most of
predictions appear to be systematically affected by a large {\it
 posetive} bias [for the binding free energy (BFE)] of the order of 3 kcal/mol. Paraquat (G13) is a
notable exception to this trend as the binding energy is in this
case {\it understimated} by less than 2 kcal/mol. However, this outcome
could be due to a cancellation effect, with the prediction being
actually much lower than the experimental counterpart when the former
is corrected for the systematic and mean signed error.  

\begin{table}[h]
  \begin{center}
  \begin{tabular}{lrrrrrrr}
    \hline\hline
Est.   &      $R_{xy}$   &  $a$ & $b$   &  MUE  & $\tau$&  MSE & npt
\\ \hline
Pred    &    0.68 &   1.12 &   1.87 &   3.07 &   0.42 &     -2.77 &  13\\
Gauss   &    0.70 &   1.25 &  -0.10 &   2.29 &   0.29 &     -1.86 &  13\\
EM(3)   &    0.67 &   1.11 &   2.18 &   3.27 &   0.46 &     -3.01 &  13\\
Jar+B   &    0.69 &   1.19 &   1.66 &   3.39 &   0.38 &     -3.14 &  13\\ \hline
Pred    &   0.76  &   0.93 &   3.37 &   3.10 &   0.50 &     -2.88  &  18\\
Gauss   &   0.76  &   0.99 &   2.19 &   2.45 &   0.50 &     -2.14  &  18\\
EM(3)   &   0.77  &   0.95 &   3.49 &   3.27 &   0.54 &     -3.09  &  18\\
Jar+B   &   0.77  &   0.98 &   3.33 &   3.38 &   0.48 &     -3.19  &  18\\\hline
Pred    &   0.93  &  0.84  &  4.27  &  3.18  &  0.80  &    -3.18 &   5\\
Gauss   &   0.92  &  0.88  &  3.70  &  2.88  &  0.80  &    -2.88 &   5\\
EM(3)   &   0.91  &  0.86  &  4.24  &  3.28  &  0.80  &    -3.28 &   5\\
Jar+B   &   0.92  &  0.87  &  4.25  &  3.34  &  0.80  &    -3.34 &   5\\
\hline \hline
  \end{tabular}
  \end{center}
\caption{Precision and accuracy metrics (see text) for the vDSSB
 estimates of Table \ref{tab:dgs}.}
  \label{tab:metr}
\end{table}
\noindent
This molecule was in fact
included also in the SAMPL7 challenge for Isaac's
TrimerTrip\cite{Amezcua2021}, and in our SAMPL7 submission paraquat
was an outlier, with a strongly {\it underestimated} dissociation free
energy, possibly due to the neglect of polarization in the bound
state.\cite{procsampl7} To the light of the structural similarities of
WP6 and the TrimerTrip, and as we used exactly the same force field
for paraquat in both challenges, we would have expected G13 to be an outlier
too in the SAMPL9 challenge.  As a matter of fact, if we remove G13
(whose binding affinity was actually known before the submission),
the Pearson correlation improves to 0.78 and the $\tau$ Kendall to
0.56 while the MUE becomes virtually identical to the MSE ($\simeq
3.1$ kcal/mol, see Table \ref{tab:ranked} further on).

\section{Discussion}
In Table \ref{tab:subs} we succintly describe  the five ranked submssions to
the SAMPL9 challenge
\begin{table}[h]
  \begin{tabular} {ccccccc}
    \hline\hline
    Sub. & Method & FF & Restr. & Charge corr. & Neutr.\\  \hline
    Voelz & ALCH & OpenFF2.0/AM1BCC & COM(strong)  &  ?   &  12 Na+ \\
    vDSSB  & ALCH  & GAFF2/AM1BCC & COM (weak) & PME &  UBP \\
    Ponder &  ALCH  & AMOEBA & COM & PME  & 12 Na+   \\
    He   &  PB/SA  & GAFF2/ABCG2 &  n/a &  ?  & 12 Na+/PB  \\
    Serrilon & ML &  n/a & n/a & n/a & n/a  \\
    \hline
  \end{tabular}
  \caption{Methodological information on the SAMPL9 ranked submissions.  ALCH, alchemical; ML,
    Machine Learning; PB/SA, Poisson Boltzmann, implicit
    solvent; FF, force field; Restr., host-guest restraint type;
    Charge corr., correction for charged ligands (finite size); Neutr.,
  Neutralization precedure for the bound state; UBP, uniform
  background plasma.}
  \label{tab:subs}
\end{table}
The Serrilon submission was done using machine learning methods based
on a neural network. The He prediction was based on an MM-PB/SA
approach (implicit solvent) on conformations generated by molecular
dynamics with explicit solvent using the GAFF2 force field.  We then
count three MD-based submissions, using a common alchemical approach
each with a different force field, namely the fixed-charge
GAFF2/AM1-BCC\cite{gaff2} (vDSSB, the present submission), 
OpenFF2.0\cite{openff} (Voltz) and the polarizable force field AMOEBA\cite{amoeba} (Ponder).
The Ponder and Voelz calculations were done using the traditional
free energy perturbation approach with equilibrium simulations for
each intermediate alchemical state along the $\lambda$-stratification.
No enhanced sampling was used in Ponder while in Voelz $\lambda$-hopping
enforced via a serial generalized ensemble (SGE) approach\cite{Berg91,chelli10,Wang2001}
with adaptive weights was used (termed ``Expanded
Ensemble'').  In $\lambda$-hopping, states can exchange $\lambda$
values so that an SGE trajectory can visit all $\lambda$ intermediate
states. However, no scaling of the potential energy barriers is
enforced, as in e.g. FEP+\cite{Wang2015} or in the end-states of
vDSSB.\cite{vdssb} The effectiveness of $\lambda$-exchanges via HREM
was recently assessed in Ref. \cite{Gapsys2021} where ``HREX
enhancement [did] not appear to give a significant boost to the FEP
accuracy''.

\begin{figure}[h]
  \begin{center}
  \includegraphics[scale=0.35]{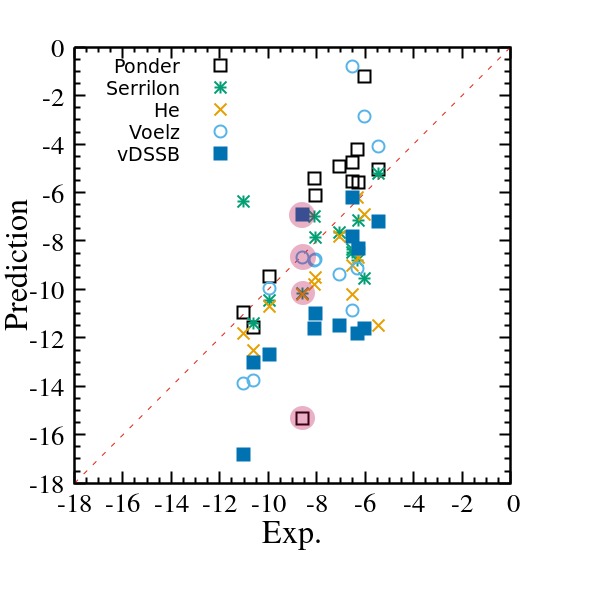}
  \end{center}
  \caption{Correlation diagrams of experimental and calculated binding
    free energy (units of kcal/mol) for the five ranked submissions. The
    red-highlighted symbols refers to G13 (paraquat).}
    \label{fig:corr_ranked}
\end{figure}

At variance with vDSSB where a uniform neutralizing
background was adopted, the MD-based Voelz and Ponder used 12 Na$^+$
counterions in the bound state.  In Table \ref{tab:ranked}, we report
results of the ranked submissions for the SAMPL9 challenge using as
metrics the Pearson correlation coefficient $R_{xy}$, the slope $a$
and intercept $b$ of the best fitting line, the Kendall ranking
coefficient $\tau$, and the mean unsigned and signed errors MUE, MSE.
\begin{table}[h]
\begin{center}
  \begin{tabular}{lllllll}
    \hline\hline
   Submission         &           $R_{xy}$  &      $a$  &      $b$   &  $\tau$  &   MUE  &  MSE\\\hline
   Serillon           &         0.38  &  0.36 &  -5.58  &  0.18  &  1.56  &  0.61\\
   He                 &         0.63  &  0.65 &  -4.61  &  0.53  &  1.89  &  1.87\\
   Ponder             &         0.76  &  1.58 &   5.28  &  0.59  &  1.98  & -0.79\\
   Voltz              &         0.72  &  1.51 &   3.25  &  0.51  &  2.27  &  0.68\\
   vDSSB              &         0.68  &  1.12 &  -1.87  &  0.42  &  3.07  &  2.77\\\hline
   Serillon $^{a}$          &         0.35  &  0.32 &  -5.73  &  0.06   &  1.56  & 0.53  \\   
   He       $^{a}$          &         0.63  &  0.64 &  -4.62  &  0.48   &  1.92  & 1.90  \\                       
   Ponder   $^{a}$          &         0.90  &  1.42 &   4.61  &  0.61   &  1.58  & -1.42  \\                       
   Voelz    $^{a}$          &         0.73  &  1.53 &   3.35  &  0.58   &  2.45  & 0.73  \\                       
   vDSSB$^{a}$              &         0.78  &  1.22 &  -1.44  &  0.56   & 3.19   & 3.14\\\hline\hline
   \multicolumn{7}{l}{$^{a}$ Metrics computed by eliminating G13 (see
     text). } 
  \end{tabular}
  \caption{Results for the SAMPL9 ranked submissions; Metrics has been. MUE, MSE, and b are given
  in kcal/mol.}
  \label{tab:ranked}
  \end{center}
\end{table}
As far as correlation is concerned, the three MD-alchemy submissions
perform decently, with the Ponder set featuring the best prediction
for both $R_{xy}$ and $\tau$ and with vDSSB yielding the best fitting
line. Ponder is still the best correlated prediction ($R_{xy}=0.90$)
if the G13 guest, with known BFE at the time of the
challenge, is removed from the set. The AMOEBA polarizable force field
used in the Ponder submission already performed quite well in the
SAMPL7 challenge on the Isaac's TrimerTrip\cite{Amezcua2021}, hence
confirming the importance of charge reorganization/polarization in
highly polar or charged molecules such as CBn-like compounds or WP6.
While the MUE is in the range of 2:3 kcal/mol for Ponder, Voelz and
vDSSB, the MSE is positive and large {\it only} for our prediction,
with the majority of the dissociation free energies significantly
overestimated.
This fact can also be visually appreciated from Figure
\ref{fig:corr_ranked}, where the correlation diagrams for the five ranked
submissions are reported. Incidentally,  we note that the BFE of G13/paraquat
(red highlighted symbols) in Serillon (-8.68 kcal/mol), Voelz and He
(10.2 kcal/mol) is in accord either with the experimental value
using fluorimetric assay reported in Ref. \cite{Yu2012}  or
with the ITC-determined association constant provided in Ref. \cite{Nicolas2017}.
A systematic overestimation of our vDSSB dissociation free
energies was expected, since, before the submission deadline, we found
excellent correlation for the five compounds G14-G18 with known
$\Delta G_d^0$ but we also found a significant mean MSE (see Table
\ref{tab:metr}).
\noindent 
In the original submission text file\cite{SAMPL9t} we
in fact wrote: ''given the negative [for dissociation free energy] MSE
found in the pre-assessment, we expect a consistent overestimation (in
the order of 2-3 kcal/mol) of the submitted BFE, possibly due the
presence of protonated WP6 species at ph=7.4''. However, such striking
discrepancy cannot be attributed to the systematic contribution of
complexes with WP6 in different protonation states as the experimentally
determined weight of the WP6$^{-12}$ species at pH=7.4 is found to
be close to unity. Besides, the large MSE is not observed in the other
two alchemy-based predictions.  The large MSE in our case is probably
caused by the approach used in the neutralization of the MD box in the
bound state. Both in Voelz and Ponder MD-based submissions,
neutralization of the 12 negative charges of the host was achieved by
introducing an equivalent number of Na$^+$ counterions.  In our
prediction, no counterions were used and neutralization was imposed
naturally by the PME approach by way a uniform background plasma with
constant charge density $\rho= 12/V_{\rm box}$, using the finite size
correction (Eq. \ref{eq:fs}) for charged guests on the computed
$\Delta G_d^0$. By analyzing the HREM trajectories of the bound state,
we found that the WP6 toroidal cavity allocating the guest molecule in
most of the cases remains completely devoid of water molecules, hence
exhibiting a lower mean dielectric constant than that of the
surrounding bulk. In Ref. \cite{Hub2014}, Hub et {\it al.} showed
that, when using a uniform background neutralizing distribution in the
context of PBC/PME, artifacts may arise in particular in systems with
an inhomogeneous dielectric constant. More in detail, the authors
demonstrated that the background charge artificially cause the charged
guest to {\it favor the lower dielectric environment}. The authors
provided a simplified correction, based on the Poisson-Bolztamnn
model, for the free energy difference in the two dielectrics (bulk
water and low dielectric environment). For spherical geometry, the
correction reads
\begin{equation}
  \Delta V = -2\pi  \frac{q_{\rm guest} \rho_{\rm BG} r_0^2}{\epsilon_l}
  \label{eq:corr}
\end{equation}
where $\epsilon_l$ is the permittivity of the low-dielectric, $r_0$ is the
radius of the sphere, and $\rho_{\rm BG}$ is the uniform background
charge density.  The approximated correction Eq. \ref{eq:corr} for the
spherical geometry cannot be lightly used for the WP6 toroidal cavity
with an unknown permittivity, possibly dependent on the guest size and
topology.  However, we can use Eq. \ref{eq:corr} to have an order of magnitude of this
correction. If we set $\epsilon_l$ in the range 2:4, and
$r_0=3.5:4~\AA$ (the effective radius of the WP6 cavity), we find
corrections in the range -1.8:3.0 kcal/mol. For the Voelz and Ponder
prediction sets these corrections are one order of magnitude smaller
as their $\rho_{\rm BG}$ includes {\it only} the neutralizing charge
for the guest molecule being the WP6$^{-12}$ anion neutralized by the
twelve Na$^+$ cations.  

A Hub-like correction for inhomogeneous dielectric environments in the
WP6 complexation may indeed explain why our dissociation free
energies, obtained with a neutralizing uniform $\rho$ of 12 positive
charges, are systematically overestimated while in Voelz and Ponder
submissions, where the neutralization was achieved by adding twelve
explicit cations, positive and negative MSE below 1 kcal/mol were
found.  Moreover, the correction Eq. \ref{eq:corr} would make the
dicationic G13 guest an outlier with an MSE in the range -4:-5
kcal/mol. If we eliminate G13 from the set, such outcome is consistent
with the result obtained for G13 in the case of the SAMPL7 TrimerTrip
(characterized by a quasi-toroidal cavity of similar size as WP6) where
the dissociation free energy was underestimated by more than 5
kcal/mol using the same force field and a strictly related MD-based
alchemical approach.\cite{procsampl7} On the other hand, it should be
noted that the vDSSB dissociation free energy of the {\it chiral}
dicationic G12 guest even corrected using Eq. \ref{eq:corr}, would still be
significantly overestimated. In this case, a possible explanation for
the discrepancy can be that, as the experimental binding constant
refers to the {\it racemic} mixture\cite{sampl9exp}, the affinities of
the two bulky stereoisomers are disparate and the effective binding
constant would be dominated by the {\it weakest} binder.

Verifying the hypothesis for the consistent overestimation of the
dissociation free energy stemming from the uniform background plasma is
straightforward but costly. We would need to repeat all HREM
and NE calculations for the bound state resizing our MD box using a
sidelength of $\simeq$ 60 \AA~to match the experimental Na$^+$
concentration of 0.14 M\cite{sampl9exp}, increasing the volume (and
the cost) by a factor of four. Besides, convergence of a solution
containing diffusing cations and a single heavily charged host-guest
complex is certainly harder than that where neutralization is provided
by a constant uniform background. In vDSSB, canonical sampling is essential in
the end-state for the bound complex. Our HREM scaling protocol, as
stated in the Methods section, only involves intra-solute interactions
leaving unaltered the solute-solvent balance so as not to accelerate
passive diffusion. This is certainly an asset in our approach as the
impact on the residence time of the (weakly restrained) guest in the
binding pocket is limited but can be an important shortcoming when
using explicit counterions by significantly slowing down convergence.

\section{Potential pitfalls and weak points in vDSSB}

We have previously seen that the estimate type from the work
convolution for $\Delta G_{\rm vdssb}$ is based on Eq. \ref{eq:jarz},
\ref{eq:em} or on the Gaussian assumption depending on the character of
the bound and unbound work distributions as assessed by the AD test or
on the uncertainty of the EM estimate. Eq. \ref{eq:jarz} is used when
the AD tests fail and when the EM estimate on the convolution has a
large ($\ge 2.0$ kcal/mol) uncertainty. In this case Eq. \ref{eq:jarz}
is corrected for a theoretical $\sigma$ and $n$ dependent positive bias
computed from a Gaussian distribution with the same variance of the
convolution work distribution. If the latter departs significantly
from a normal distribution, the bias can be overestimated with an
impact on the accuracy of $\Delta G_{\rm vdssb}$. 

A weak point of {\it any} alchemical approach, including vDSSB, is
related the standard state correction, connected to the elusive
binding site volume\cite{procsampl7,statbind,Luo2002,Gilson1997}. In
vDSSB, where a weak COM-COM restraint is used, the {\it negative}
correction to the {\it dissociation} free energy is estimated from the
variance of the host-guest COM-COM distance distribution in the bound
state and is hence a characteristic of the host-guest complex, varying
in a range -5:-2.5 kcal/mol (see Table \ref{tab:dat}). In principle,
the binding site volume should depend on both the host and the guest
in the context of the NE alchemical theory\cite{pccp1}.  In Voelz and
Ponder FEP calculations, the volume correction is related to the force
constant of the host-guest COM-COM restraint as $V_{\rm restr}=
(\frac{2\pi RT}{k} )^{3/2}$ and, as such, is {\it identical for all
  guests}, in the assumption that in the simulation of the restrained
complex all states that are made accessible by restrained potential
have been canonically sampled for all $\lambda$ states along the
alchemical path\cite{Gilson1997,statbind} Such correction to the
dissociation free energy is equal to -3.85 in Voelz\cite{SAMPL9v}. In
Ponder, again an {\it identical} volume correction was used for all
guests using COM-COM restraints defining a flat-bottom potential
``chosen such that [these restraints] were not violated during
unrestrained simulations runs on the bound host-guest
complex.''\cite{SAMPL9p} The size of the Ponder standard state
correction was not given in Ref. \cite{SAMPL9p}.  If we remove our
guest-dependent standard state volume correction ($\Delta G_{\rm vol}$
entry in Table \ref{tab:dat}) assuming an identical correction of -4
kcal/mol for all guests as in Ponder and Voelz, agreement with
experiment luckily improves, yielding $R_{xy}=0.84$, $\tau=0.49$,
$a=1.60$, MUE=2.18 kcal/mol and MSE=1.50.

\section{Conclusion}

We have presented our results on the SAMPL9 challenge focusing on
binding of soluble carboxy-pillar[6]arene)with ammonium/diammonium
cationic guests. We adopted an alchemical technique, named virtual
double system single box, relying on an enhanced sampling of the
end-states and on the production of few hundreds of fast
nonequilibrium trajectories where the ligand  is decoupled and recoupled in
the bound and unbound state, respectively.  The resulting work
distributions were combined (convoluted) as if the two independent
processes occurred in the same box, with one ligand on the host and
the other in the far distant bulk. The dissociation free energy was
estimated from the convolution using the Jarzynski theorem, the
Gaussian assumption, or the Gaussian mixture assumption, depending on
the character of the bound and unbound work distributions.   

Correlation with experimental data is acceptable and consistent with
our previous SAMPL6 and SAMPL7 submission also based on nonequilibrium
alchemy.  If G13 (paraquat), whose experimental binding constant was
known at the time of the challenge\cite{SAMPL9t}, is removed from the
set, vDSSB is the second best-correlated submission of the challenge
(see Table \ref{tab:ranked}).  Nonetheless, when compared to the other
two MD-based SAMPL9 alchemical submissions, our prediction set for
the dissociation free energy exhibits a large mean signed error (MSE),
suggesting a systematic error in the simulation protocol. We attribute
the large MSE to the way neutralization was imposed in the bound state
containing the WP6$^{-12}$ anion and the guest. At variance with the
other MD-based submissions, where explicit counterions were
introduced, we used a uniform neutralizing background plasma. The
uniform background charge approach has been recently demonstrated to
artificially favor the binding in the lower dielectric
environment.

Overall, the challenge has confirmed that MD-based alchemical methods
tend to correlate better with experiment with respect to other
approaches (e.g. Machine Learning or Poisson-Boltzmann with
MD sampling of the end-states). However, accuracy is still falling
short in MD protocols, with a mean deviation of the order of 2 kcal/mol
even when resorting to computationally demanding polarizable force
field. In this regard, we note that the present SAMPL9 challenge (as
well as some of the past challenges) is characterized by exceptionally
critical issues from the parameterization perspective mostly related
to electrostatic interactions, such as unknown protonation state of
the highly charged host molecule, chiral specificity in binding,
singly or doubly charged guests. Besides, in most of the recent SAMPL
initiatives, all guest molecules share in general a common chemical
structure, characterized by charged substituents on a hydrophobic
scaffold while the hosts are rigid macrocyclic molecules heavily
decorated with polar or charged groups. As recently shown in
Ref. \cite{Vassetti2019}, small imperfections on the electrostatic
modeling can have a huge impact on the BFE.  Possibly,
these imperfections could be systematically enhanced when testing
MD-based approaches with ``general'' electrostatic parameterizations
in the recent SAMPL challenges. 

\backmatter

\bmhead{Supplementary information}

The bound and unbound state work distributions for all eighteen
ligands of Figure \ref{fig:sampl9} are reported in the Supporting
Information. PDB trajectory files, raw work data, and force field
parameter files are available at the general-purpose open-access
repository Zenodo (https://zenodo.org/record/5891191).

The ORAC program (v6.1) is available for download under the GPL at the
website  http://www1.chim.unifi.it/orac/

Third-party software Autodock Vina can be downloaded from the
website https://vina.scripps.edu/ 

\bmhead{Acknowledgments}
The computing resources and the related technical support used for
this work have been provided by CRESCO/ENEAGRID High Performance
Computing infrastructure and its staff. CRESCO/ENEAGRID
m,High Performance Computing infrastructure is funded by ENEA, the
Italian National Agency for New Technologies, Energy and Sustainable
Economic Development and by Italian and European research programmes
(see www.cresco.enea.it for information).


\end{document}